\documentclass{article}
\usepackage[T1]{fontenc} 
\usepackage[utf8]{inputenc} 
\usepackage{ismir,amsmath,cite,url}
\usepackage{graphicx}

\usepackage{color}

\usepackage[firstpage]{draftwatermark}
\definecolor{lightgray}{rgb}{0.9,0.9,0.9}
\definecolor{darkgray}{rgb}{0.4,0.4,0.4}
\SetWatermarkFontSize{12pt}
\SetWatermarkScale{1.1}
\SetWatermarkAngle{90}
\SetWatermarkHorCenter{202mm}
\SetWatermarkVerCenter{170mm}
\SetWatermarkColor{darkgray}
\SetWatermarkText{Late-Breaking / Demo Session Extended Abstract, ISMIR 2024 Conference}

\def\lbd{}	

\title{PyNeuralFx: A Python Package for Neural Audio Effect Modeling}

\threeauthors
  {Yen-Tung Yeh} {National Taiwan University \\ {\tt r12942179@ntu.edu.tw}}
  {Wen-Yi Hsiao} { Indepedent Researcher \\\tt s101062219@gmail.com}
  {Yi-Hsuan Yang} {National Taiwan University \\ {\tt yhyangtw@ntu.edu.tw}}

\def\authorname{Yen-Tung Yeh, Wen-Yi Hsiao, and Yi-Hsuan}

\begin{document}

\maketitle
\begin{abstract}

We present PyNeuralFx, an open-source Python toolkit designed for research on neural audio effect modeling.
The toolkit provides an intuitive framework and offers a comprehensive suite of features, including standardized implementation of  well-established model architectures, loss functions, and easy-to-use visualization tools. 
As such, it helps promote reproducibility for research on neural audio effect modeling, and enable in-depth performance comparison of different models, offering insight into the behavior and operational characteristics of models through DSP methodology. The toolkit can be found at \url{https://github.com/ytsrt66589/pyneuralfx}. 

\end{abstract}
\section{Introduction}\label{sec:introduction}

Neural audio effect modeling aims to use neural networks to simulate and replicate various audio effects typically achieved through traditional digital signal processing (DSP) techniques. Several studies have demonstrated that neural networks can achieve high-quality emulation results \cite{8683529,769f627fa4fe49569bd207f6b1d32dc3,wright2022grey,steinmetz2022efficient}. However, different works often employ varied training strategies, loss functions, and evaluation metrics, making it hard to compare models. For example, aiming to emulate compressor, the energy-to-signal ratio loss is used in \cite{wright2022grey} but the hybrid loss of mean absolute error and multi-resolution short-time Fourier transform loss is used in \cite{steinmetz2022efficient}. Little discussion is on comparing how the loss affect the performance on different types of audio effects. 

Moreover, the lack of interpretability in neural network systems hinders our understanding of their inner workings, a challenge particularly significant in neural audio effect modeling. This limitation is crucial for two reasons. First, without comprehending the actual behavior of these models, musicians face difficulties in fine-tuning sonic characteristics and addressing unexpected behaviors. Second, neural network models can implicitly conceal issues such as aliasing, which are not easily observed without insight into the internal processes of a system. Given the rising importance of the task, an easy-to-use toolkit 
is needed. 

To tackle these issues, we propose PyNeuralFx, a toolkit that offers standarized implementation of several training strategies and loss functions, making it easy to benchmark research on audio effect modeling. Furthermore, we provide  visualization tools to offer insights into the neural networks. For instance, users can visualize the harmonic response of the models to study the system's behavior. PyNeuralFx aims to standardize the approach to neural audio effect modeling, fostering reproducibility and facilitating more meaningful comparisons between different models. By offering a unified platform for experimentation and analysis, we seek to advance the field and promote a deeper understanding of neural network-based audio processing systems. 
We provide some details of the toolkit below.

\begin{figure}
 \includegraphics[width=1\columnwidth]{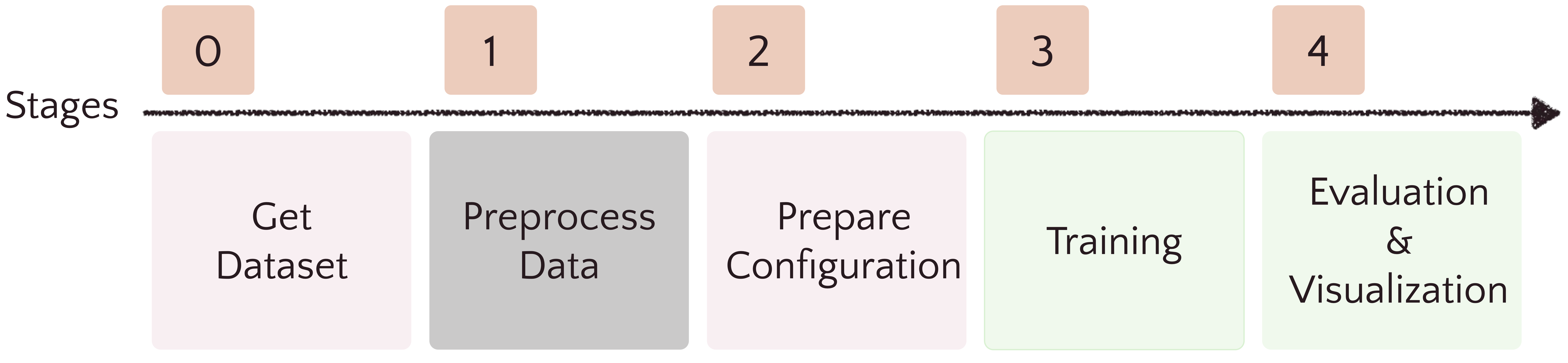}
 \caption{Workflow of the proposed PyNeuralFx toolkit.}
 \label{fig:flow}
\end{figure}

\section{Functionality}


\subsection{Model Architectures}
We implement a range of \emph{modeling backbones} and \emph{control mechanisms}. 
The former can be conceptualized as the ``effect processor,'' while the latter serves to inject specific conditions into the model. 
A user can form different combinations of modeling backbones and control mechanisms, permitting flexible and powerful audio effect modeling. 

For modeling backbones, we have both CNN and RNN based networks.
CNNs include models widely-used in  audio effect modeling, including TCNs (temporal convolutional networks) \cite{steinmetz2022efficient} and GCNs (gated convolution network) \cite{10097173,yeh24dafx, chen2024zeroshotamplifiermodelingonetomany}. RNNs emcompass the vanilla RNN, LSTM, and GRU architectures \cite{769f627fa4fe49569bd207f6b1d32dc3, yeh24dafx, 6567472, schmitz2018real}. These diverse options allow researchers to explore and compare various neural network structures within a unified framework. 

For control mechanisms, we note that in the literature, CNN backbones tend to go with either the concatenation method (Concat) \cite{8682805} or feature-wise linear modulation (FiLM) \cite{steinmetz2022efficient}, while RNNs mostly use the Concat method only \cite{769f627fa4fe49569bd207f6b1d32dc3}.
We implement these all.
Moreover, we include three novel conditioning methods for RNN proposed in our recent work \cite{yeh24dafx}, including FiLM and two \emph{hypernetwork}-based methods, called `StaticHyper' and `DynamicHyper.'
In particular, based on our previous finding of the effectiveness of hypernetwork-based conditioning for RNNs \cite{yeh24dafx}, we also implement a hypernetwork-based conditioning method for CNNs, though such a combination has not been found elsewhere in neural audio effect modeling. 


\begin{figure}
 \centerline{
 \includegraphics[width=1\columnwidth,height=0.13\textheight,keepaspectratio]{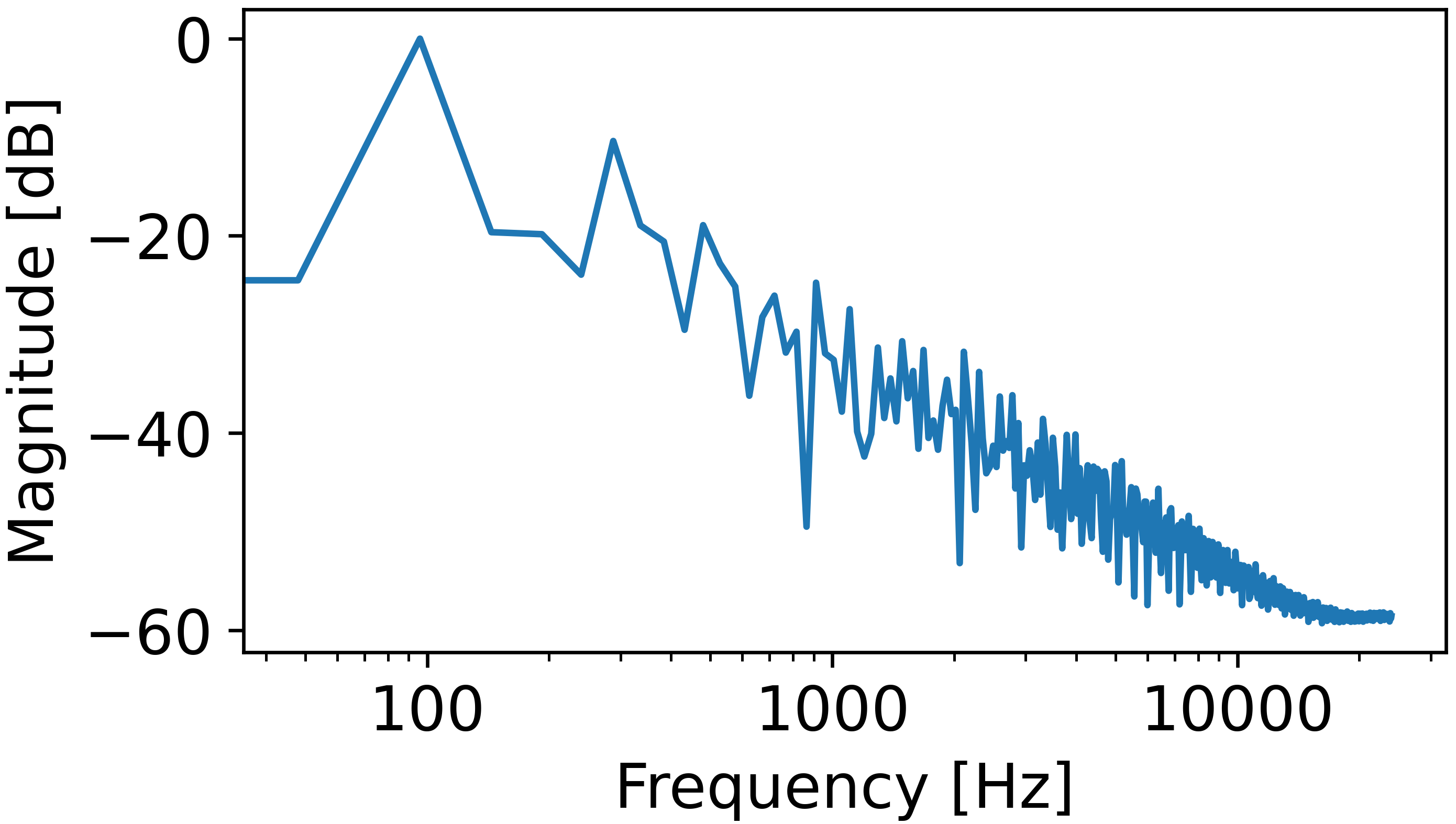}}
 \caption{Visualization of the harmonic response of a model trained on Boss OD-3 overdrive device \cite{yeh24dafx}.} 
 \label{fig:harmonic_response}
\end{figure}

\subsection{Loss Functions for Training}
\label{sec:loss_func}
We implement several existing and new loss functions used in neural audio effect modeling, such as the widely-used energy-to-signal (ESR) ratio \cite{769f627fa4fe49569bd207f6b1d32dc3}, hybrid combination of time-domain mean absolute error and frequency-domain multi-resolution short-time Fourier transform loss \cite{steinmetz2022efficient,yeh24dafx}, and the short-time Fourier transform complex loss used in another recent publication from our group \cite{chen2024zeroshotamplifiermodelingonetomany}. Moreover, we also support different pre-emphasis filter strategies proposed in \cite{9052944}, and the DC loss proposed in \cite{769f627fa4fe49569bd207f6b1d32dc3} for reducing the DC offset in the model outputs.

\subsection{Objective Metrics for Evaluation}

Most neural audio effect systems are often evaluated using the reconstruction loss alone \cite{769f627fa4fe49569bd207f6b1d32dc3, 10097173, 8682805}, possibly due to the lack of accessible libraries that offer alternative metrics.
This is limited as the reconstruction loss only paints a partial picture of the performance of a model.
We address this issue by supporting several evaluation metrics sporadically used in previous works, including the loudness error \cite{schmitz2018real}, crest factor \cite{yeh24dafx}, RMS energy \cite{yeh24dafx}, and the transient metic newly proposed in our recent work \cite{yeh24dafx}. Additionally, inspired by \cite{steinmetz2022style}, we incorporate spectral centroid as an evaluation metric. This comprehensive set of metrics allows for a multifaceted assessment of model performance, capturing various aspects of audio quality and fidelity.

\subsection{Visualization Tools}

We currently offer two types of visualization tools to aid in the analysis and interpretation of neural audio effect models. The \texttt{pyneuralfx.vis.plotting} enables direct comparison of waveforms, allowing a user to visually assess the similarities and differences between the original and processed audio signals. The \texttt{pyneuralfx.vis.sysplotting} involves sending test signals through the system to observe the corresponding responses of the neural network.

\begin{figure}
 \centerline{
 \includegraphics[width=1\columnwidth,height=0.12\textheight,keepaspectratio]{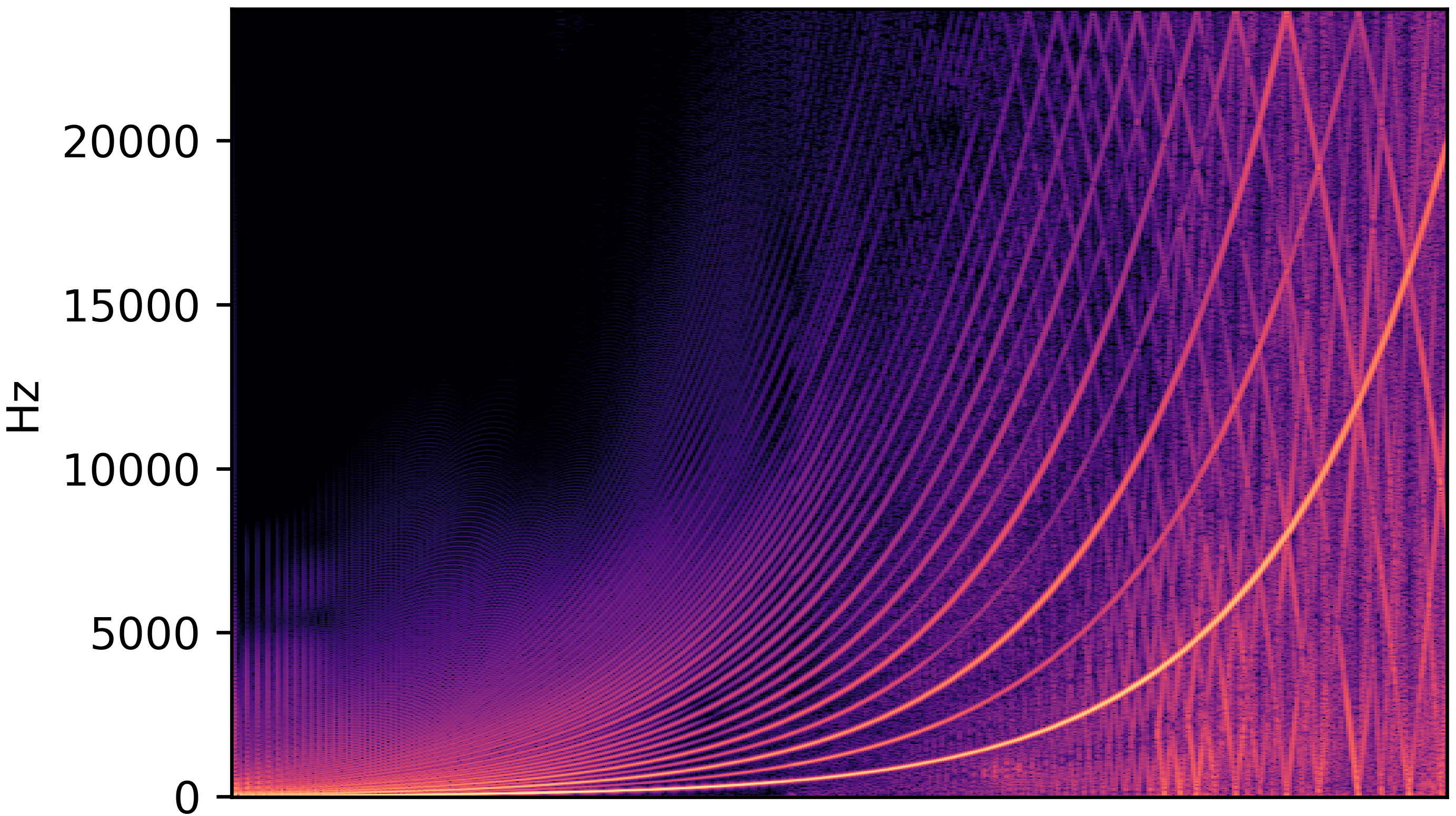}}
 \caption{Sine sweep response of a  network model, revealing the aliasing problem (i.e., high frequency folds back).}
 \label{fig:aliasing}
\end{figure}

\section{Usage flow}

We demonstrate the workflow of the PyNeuralFx, allowing researchers to standardize experiment easily. 

\textbf{Get Dataset. } To begin with, prepare a dataset either from  academic resources such as SignalTrain \cite{hawley2019signaltrain}, EGDB \cite{9747697}, or Boss OD-3 \cite{yeh24dafx}, or by creating your own custom dataset tailored to emulate your target audio effect. The choice between using existing datasets and building your own depends on the audio effects you aim to model.

\textbf{Preprocess Data. } Due to the diverse and highly customizable formats of audio datasets, providing a universal framework for processing all datasets is challenging. Our solution is to offer a standardized template that our framework supports. Researchers can then preprocess their data to match this required template. This approach ensures flexibility while maintaining compatibility with our toolkit. Consequently, data preparation might be the only stage where users need to write custom code, streamlining the overall workflow and maximizing the usability of the toolkit across various research scenarios.

\textbf{Prepare Configuration. } PyNeuralFx runs each experiments based on the configuration .yml files. Each file will record every training details to improve the reproducibility across each experiments.

\textbf{Training, Evaluation \& Visualization. } After preparing all materials, users can run the training process via PyNeuralFx. When the training is finished, users can evaluate the result through several prepared metrics and visualize the system's behavior. Example of the harmonic response visualization is shown in Figure \ref{fig:harmonic_response}, and the aliasing problem is shown in Figure \ref{fig:aliasing}.

\section{Conclusion}

In this paper, we have presented PyNeuralFx, a toolkit designed to support research on neural audio effect modeling. 
Part of the codebase of PyNeuralFx is integrated from those of our two recent papers \cite{yeh24dafx, chen2024zeroshotamplifiermodelingonetomany}, suggesting the relevance of the toolkit.
For future work, we aim to expand the capabilities of PyNeuralFx by providing pre-trained models for various audio effects, 
incorporating new architectural advancements (e.g., \cite{yin24ssm}), and implementing additional evaluation metrics and loss functions  (e.g., GAN-based ones \cite{wright2023adversarial,chen24dafx}). 
We are also open to contributions and joint work from the community.

\bibliography{ISMIRtemplate}

%
%
%
%
%

\end{document}